\documentclass[twocolumn,aps,floatfix,showpacs,superscriptaddress,amsmath,amssymb,intlimits]{revtex4}
\newcommand{\be}{\begin{equation}}
\newcommand{\ee}{\end{equation}}
\newcommand{\bea}{\begin{eqnarray}}
\newcommand{\eea}{\end{eqnarray}}

\usepackage{graphicx}
\usepackage{bm}
\begin{document}

\begin{abstract}


We study the role played by the magnetic dipole interaction in an
atomic interferometer based on an alkali Bose-Einstein condensate
with tunable scattering length. We tune the s-wave interaction to
zero using a magnetic Feshbach resonance and measure the
decoherence of the interferometer induced by the weak residual
interaction between the magnetic dipoles of the atoms. We prove
that with a proper choice of the scattering length it is possible
to compensate for the dipolar interaction and extend the coherence
time of the interferometer. We put in evidence the anisotropic
character of the dipolar interaction by working with two different
experimental configurations for which the minima of decoherence
are achieved for a positive and a negative value of the scattering
length, respectively. Our results are supported by a theoretical
model we develop. This model indicates that the magnetic dipole
interaction should not represent a serious source of decoherence
in atom interferometers based on Bose-Einstein condensates.

\end{abstract}

\title{Magnetic dipolar interaction in an atomic Bose Einstein condensate interferometer}
\author{M. Fattori}
\affiliation{LENS and Dipartimento di Fisica, Universit\`a di
Firenze,
  and INFM-CNR\\  Via Nello Carrara 1, 50019 Sesto Fiorentino, Italy }
\affiliation{Museo Storico della Fisica e Centro Studi e Ricerche
'Enrico Fermi' ,Compendio del Viminale, 00184 Roma, Italy }

\author{G. Roati}
\affiliation{LENS and Dipartimento di Fisica, Universit\`a di
Firenze,
  and INFM-CNR\\  Via Nello Carrara 1, 50019 Sesto Fiorentino, Italy }
\affiliation{INFN, Sezione di Firenze, Via Sansone 1, 50019 Sesto
Fiorentino, Italy }

\author{B. Deissler}
\affiliation{LENS and Dipartimento di Fisica, Universit\`a di
Firenze,
  and INFM-CNR\\  Via Nello Carrara 1, 50019 Sesto Fiorentino, Italy }

\author{C. D'Errico}
\affiliation{LENS and Dipartimento di Fisica, Universit\`a di
Firenze,
  and INFM-CNR\\  Via Nello Carrara 1, 50019 Sesto Fiorentino, Italy }
\affiliation{INFN, Sezione di Firenze, Via Sansone 1, 50019 Sesto
Fiorentino, Italy }

\author{M. Zaccanti}
\affiliation{LENS and Dipartimento di Fisica, Universit\`a di
Firenze,
  and INFM-CNR\\  Via Nello Carrara 1, 50019 Sesto Fiorentino, Italy }

\author{M.~Jona-Lasinio}
\affiliation{LENS and Dipartimento di Fisica, Universit\`a di
Firenze,
  and INFM-CNR\\  Via Nello Carrara 1, 50019 Sesto Fiorentino, Italy }

\author{L. Santos}
\affiliation{Institut f\"ur Theoretische Physik, Leibniz
Universit\"at, D-30167 Hannover, Germany}

\author{M. Inguscio}
\affiliation{LENS and Dipartimento di Fisica, Universit\`a di
Firenze,
  and INFM-CNR\\  Via Nello Carrara 1, 50019 Sesto Fiorentino, Italy }
\affiliation{INFN, Sezione di Firenze, Via Sansone 1, 50019 Sesto
Fiorentino, Italy }

\author{G. Modugno}
\affiliation{LENS and Dipartimento di Fisica, Universit\`a di
Firenze,
  and INFM-CNR\\  Via Nello Carrara 1, 50019 Sesto Fiorentino, Italy }
\affiliation{INFN, Sezione di Firenze, Via Sansone 1, 50019 Sesto
Fiorentino, Italy }

\maketitle

Atom-atom interactions represent a fundamental limit to the
performance of atomic Bose Einstein condensate (BEC)
interferometers \cite{CastinDalib,Shin,KSqueezing,
KasevichSqueezing}. Atomic collisions lead to density-dependent
shifts in the interferometric signal, severely compromising its
visibility. In two recent works \cite{Fattori, Naegerl}, the
possibility to strongly reduce the interaction-induced decoherence
in a trapped BEC interferometer has been demonstrated by tuning
the s-wave scattering length $a$ almost to zero via a magnetic
Feshbach resonance. The tunability of $a$ by magnetic means is
possible for atoms with a non-vanishing magnetic dipole moment.
Therefore, once the s-wave contact interaction is canceled by
applying a proper external magnetic field, the magnetic
dipole-dipole interaction (MDI) between the atoms remains as a
possible source of decoherence for the interferometer. This aspect
has been pointed out in \cite{Fattori, Naegerl}, but both a
theoretical analysis and an experimental study of the problem are
still missing. The MDI generally does not play a role in
experiments with ultra-cold quantum degenerate alkali atoms, where
the small magnetic dipole moment $\mu$ is on the order of the Bohr
magneton $\mu_B$ and leads to a dipolar interaction energy $E_{d}
< 0.01 \cdot E_s$, with $E_s$ the s-wave contact interaction
energy. So far, studies of the MDI in an ultra-cold gas have been
possible mainly with Cr atoms \cite{Fattori2}, characterized by a
large magnetic dipole moment $\mu = 6 \mu_B$ that leads to
interaction energies $E_{d} $ 36 times larger than for alkali
atoms. Evidence of MDI in a spinorial alkali BEC has been only
recently reported in \cite{Stamper}.

In this Letter we study the role played by the MDI in an
interferometer where a BEC with weak tunable contact interaction
is implemented \cite{Fattori}. We get evidence of the effect of
dipolar interaction on the dephasing of the interferometric
signal. The MDI is anisotropic and therefore the sign of its
contribution to the interaction energy depends on the geometry of
the system. We study in particular two different geometries for
which the minimum of decoherence occurs for two different values
of the contact interaction, one positive and the other negative.
We develop a model that confirms that the minimum of the
decoherence is obtained when the contact interaction partially
compensates the MDI. The model indicates that the unavoidable MDI
should not represent a seriously limiting source of decoherence in
BEC-based atom interferometers.

For our studies we implement a Bloch oscillation interferometer
\cite{Salomon,Kasevich}. A trapped BEC of $^{39}$K atoms
\cite{k39} is loaded in a deep 1D optical lattice (OL) and an
external force $F_{ext}$ along the lattice drives Bloch
oscillations. We work with atoms in the absolute ground state
$|F=1, M_F=1 \rangle$ where the magnetic dipole moment $\vec \mu$
is parallel to the external magnetic field $\vec B$  that is
applied to access Feshbach resonances. We can align the OL either
along or orthogonal to $\vec B$. Changing $|\vec B|$ around 350 G
it is possible to finely tune $a$ around a zero crossing
\cite{k39, FSK39}. The scattering length can be controlled down to
the level of 0.06 $a_0$, where the MDI described by the two body
potential
\begin{equation} \label{dipolar pot}
V_d(\vec r)=-\frac{\mu_0|\vec \mu|^2}{4\pi}\left(\frac{3(\hat \mu
\cdot \hat r)^2 - 1}{r^3}\right)
\end{equation}
comes into play. In Eq. \eqref{dipolar pot} $\hat \mu = \vec
\mu/|\vec \mu|$ and $r = |\vec r|$ is the distance between the two
interacting dipoles. Note that the effective dipole moment of
$^{39}$K atoms at 350G is $\mu$= 0.95 $\mu_B$ \cite{DipoloEff}.
Due to a quasi 2D geometry of the optical potential in each
lattice site, the on-site MDI depends on the orientation of $|\vec
\mu|$ with respect to the OL. When the dipoles are parallel to the
OL (see Fig. 1a)), their mutual interaction within each site is
mainly repulsive. A weaker but not negligible attractive
contribution comes from distant sites due to the long range
character of the MDI. The non-uniform population over the OL leads
to a non homogeneous positive mean field shift causing dephasing
of the Bloch oscillations \cite{Fattori, Korsch}. A proper
negative value of $a$ reduces and flattens the interaction mean
field shift, increasing the coherence time of the interferometer.
In the other configuration, for dipoles orthogonal to the OL (see
Fig. 1b), the on site MDI is mainly attractive, the inter sites
MDI is slightly repulsive and a proper positive value of $a$
minimizes the decoherence.
\begin{figure}
\includegraphics[width=\columnwidth]{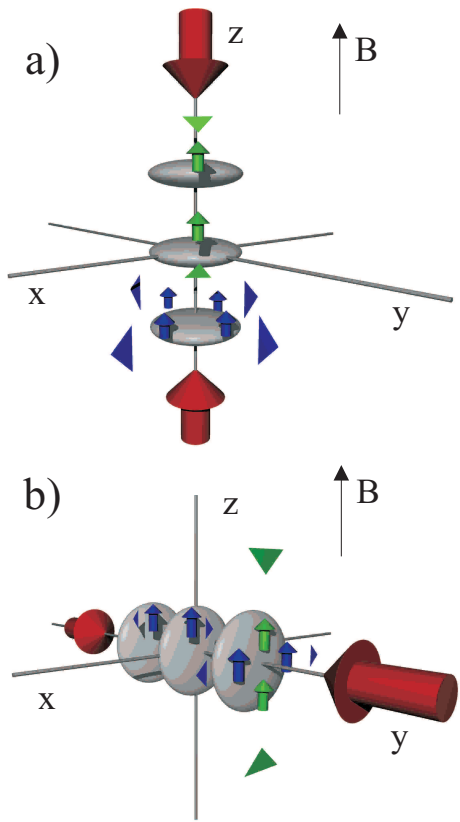}
\caption{(Color online) Schematic representation of the character
of the magnetic dipolar interaction in the two different
experimental configurations. a) For dipoles parallel to the
lattice direction, the on-site MDI is repulsive, while the weaker
inter-site MDI is attractive. b) For dipoles orthogonal to the
lattice direction, the on-site MDI is attractive, while the
inter-site MDI is repulsive.}
\end{figure}

In a Bloch oscillation interferometer decoherence manifests itself
in a linear increase of the square root of the variance of the
atomic momentum distribution as a function of the Bloch
oscillations time $t_{osc}$. It is possible to determine a rate of
decoherence from a single measurement of the normalized momentum
variance taken at large $t_{osc}$ \cite{Fattori}. Experimentally
we measure the momentum distribution by releasing the BEC from the
OL and by performing absorption imaging of the atomic density
after an expansion of 12 ms.

The experimental parameters chosen for the measurement of the
decoherence rate in the two configurations are listed below. For
the OL $\parallel \vec B$, we implement a BEC of $4 \times 10^4$
atoms initially trapped in a harmonic trap with $(\nu_x, \nu_y,
\nu_z)=(76, 44, 43)$Hz. Before starting Bloch oscillations $a$ is
adiabatically tuned to 3 $a_0$. The OL has $\nu_x = \nu_y$= 44Hz
and a depth $sE_r$, where $s=6$, $E_r = \hbar^2 k_L^2/2m$ is the
recoil energy, $k_L = 2\pi/\lambda$ is the laser wavevector
($\lambda=1032$ nm) and $m$ is the atomic mass. In this
configuration $F_{ext}$ is the gravitational force. Right after
the start of Bloch oscillations, triggered by the switching off of
the harmonic trap, $a$ is tuned to a final value around the
zero-crossing by tuning the magnetic field. The minimum of
decoherence is found at $B_{\perp}$=$(349.94 \pm 0.02 \pm 0.1)$ G
(Fig.2, circles) (the first uncertainty is statistical, the second
one is systematic and comes from the uncertainty in the
calibration of the external magnetic field). For the OL $\perp
\vec B$, Bloch oscillations are driven by a spurious magnetic
field gradient generated by the Feshbach coils and the resultant
force on the atoms is six times smaller than gravity. For this
measurement we use initial trapping frequencies of (99, 45, 109)
Hz, a radial lattice confinement $\nu_x = \nu_z$ = 99 Hz,
$t_{osc}$= 300 ms, $\lambda=1064$ nm and an average atom number of
$2.5 \times 10^4$. Results are shown in Fig. 2 (squares). The
minimum of decoherence occurs for a different value of $|\vec B|$,
i.e. $B_{\parallel}$=$(350.59 \pm 0.02 \pm 0.1)$ G. Our knowledge
of the zero-crossing location ($B_{zc}=350.4 \pm 0.4$ G) is based
on Feshbach spectroscopy analysis \cite{FSK39}. Despite this
relatively large uncertainty, one notes that the two minima of
decoherence sit on the left and on the right of 350.4 G
respectively, in accordance with the qualitative explanation
presented above.

\begin{figure}
\includegraphics[width=\columnwidth]{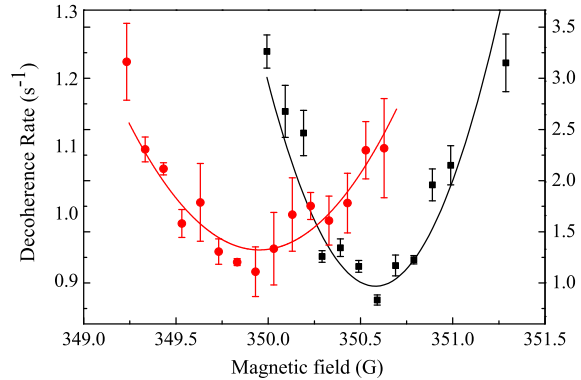}
\caption{(Color online) Decoherence rate of the interferometer as
a function of the external magnetic field applied during Bloch
oscillations for a lattice parallel (circles, left vertical scale)
and orthogonal (squares, right vertical scale) to $\vec B$.}
\end{figure}

For a ore quantitative analysis of our findings we have developed
a simple theoretical model to describe Bloch oscillations in the
presence of the MDI and a weak contact interaction. At
sufficiently low interaction strength our system can be described
by a non-local non-linear Schr\"odinger  equation (NLSE) of the
form
\begin{multline} \label{NLSE}
i\hbar\frac{\partial}{\partial t}\Psi(\vec r,t)= \left[
-\frac{\hbar^2}{2m}\nabla^2+V_L(z)+V_\perp(\rho) - F_{ext} z +
\right.
\\ \left. + g|\Psi(\vec r,t)|^2 +\int d\vec r' V_d(\vec r-\vec
r')|\Psi(\vec r',t)|^2 \right] \Psi(\vec r,t)
\end{multline}
where $V_L(z) = s E_r \sin^2(k_L z)$ is the lattice potential,
$V_\perp(\rho) = m\omega_\perp^2 \rho^2 / 2$ describes the
transversal harmonic trapping confinement and $g=4\pi\hbar^2a/m$.
In order to study the two experimental configurations, we fix for
simplicity the direction of the lattice along $\hat z$ and change
the orientation $\hat \mu$ of the dipoles. When the lattice  depth
is sufficiently large, we can implement a tight-binding model. In
particular we consider situations where the total interaction
energy is much smaller than $s E_r$ and $\hbar\omega_\perp$.
Therefore we write $\Psi(\vec r,t)= \sqrt{N} \phi(\rho)
\sum_{j}\psi_j (t) w(z-z_j), $ where $\phi(\rho)=e^{-\rho^2/2
l_\perp^2}/\sqrt{\pi}l_\perp$ is the transversal ground state,
with $l_\perp=\sqrt{\hbar/m\omega_\perp}$, and $w(z-z_j)$ is the
Wannier function associated with the lowest energy band at the
$j$th lattice site located at $bj$, $b=\pi/k_L$ being the lattice
step. For sufficiently deep lattices the Wannier functions are
well represented by Gaussians of the form $w(z)=e^{-z^2/2
l^2}/\pi^{1/4}\sqrt{l}$, with $b/l=\pi s^{1/4}$. Plugging the
tight binding ansatz for $\Psi(\vec r,t)$ into eq. \eqref{NLSE}
and integrating out the spatial coordinates we obtain the discrete
NLSE

\begin{multline} \label{DNLSE}
i\hbar\frac{\partial}{\partial t}\psi_j=-J(\psi_{j+1}+\psi_{j-1})+
\Delta j\psi_j + N U^c(a) |\psi_j|^2\psi_j + \\ + N U^{dd}_{j, j}
|\psi_j|^2\psi_j +N\sum_{\delta\neq 0}U^{dd}_{j, j+\delta}
|\psi_{j+\delta}|^2\psi_j
\end{multline}
where the five terms on the right are consecutively the tunneling
energy, the potential energy due to the external force, the
on-site contact interaction term, the on-site MDI term and the
inter-site MDI term. In particular
$J=\frac{4}{\sqrt{\pi}}s^{3/4}e^{-2\sqrt{s}}E_r$, $\Delta=-F_{ext}
b$,
\begin{equation}
U^c(a)=\frac{4\pi\hbar^2}{m}\frac{a}{(2\pi)^{3/2}l_\perp^2 l}
\end{equation}
\begin{equation} \label{dd-energy-onsite}
U^{dd}_{j,j} =\xi \frac{\mu_0\mu^2}{4\pi} \frac{1}{l_\perp^3 c^3}
\sqrt{\frac{2}{\pi}}\left[\frac{c(3-c^2)}{3\sqrt{1-c^2}}-\arcsin(c)\right]
\end{equation}
where $c = \sqrt{1-l^2/l_\perp^2}$ and where $\xi=(3(\hat \mu
\cdot \hat z)^2-1)/2$ is a geometric factor taking into account
the orientation of the dipoles $\hat \mu$ with respect to the
lattice direction $\hat z$, and
\begin{equation} \label{dd-energy}
U^{dd}_{j, j+\delta} =\xi \frac{\mu_0\mu^2}{4\pi}
\frac{1}{3l_\perp^3} \sqrt{\frac{2}{\pi}} F\left(c, \frac{\delta
b}{l_\perp}\right)
\end{equation}
where
\begin{equation}
F(u,v) = \int_0^1 ds\, \frac{3s^2-1}{(1-u^2s^2)^{3/2}}
\left(1-\frac{v^2s^2}{1-u^2s^2}\right)
e^{-\frac{v^2s^2}{2(1-u^2s^2)}}
\end{equation}
Note that Eq.\eqref{dd-energy-onsite} can be obtained by setting
$\delta=0$ in Eq. \eqref{dd-energy}.

The on-site dipole-dipole interaction energy may be re-absorbed in
the contact term by defining an effective scattering length
$a_{eff}$ such that $U^c(a)+U^{dd}_{j, j}=U^c(a_{eff})$. As a
consequence, the on-site interaction energy does not vanish at
$a=0$, but at a finite value $\bar a$ such that $a_{eff}=0$. By
equating the contact and on-site dipole-dipole interaction energy
we obtain
\begin{equation} \label{as-bar}
\bar a = -\xi \frac{\mu_0\mu^2}{4\pi}
\frac{m}{\hbar^2}\frac{\sqrt{1-c^2}}{c^3}
\left[\frac{c(3-c^2)}{3\sqrt{1-c^2}}-\arcsin(c)\right]
\end{equation}

\begin{figure}
\includegraphics[width=\columnwidth]{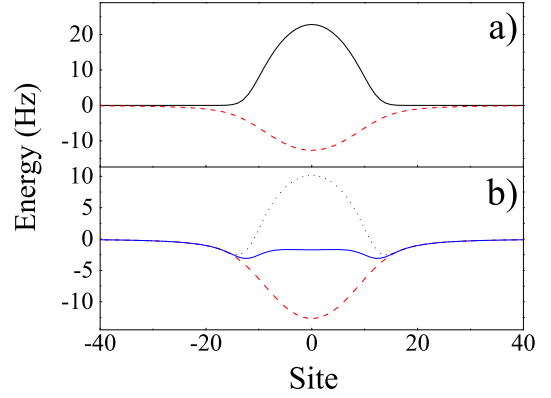}
\caption{(Color online) a) On-site dipolar interaction energy
(solid line) and inter-site dipolar interaction energy (dashed
line) as a function of the lattice site $j$ for the OL $\parallel
\vec B$ configuration. b) Total interaction energy for $a = 0$
(dotted), $a=-0.52$ $a_0$ (dashed), i.e. the value that cancels
the on-site MDI and for $a= -0.32$ $a_0$(solid), i.e. the value
that minimizes the decoherence.}
\end{figure}

In the case of $^{39}$K we have $m \mu_0\mu^2/ (4\pi\hbar^2) =
0.85$ $a_0$. Using the parameters of our experimental setup we
obtain $\bar a = -0.52$ $a_0$ for OL $\parallel \vec B$ ($\xi =
1$) and $\bar a = 0.24$ $a_0$ for OL $\perp \vec B$ ($\xi =
-1/2$). It is then clear that, if we neglect the inter-site MDI,
the minima of decoherence for the two lattice configurations would
be separated by (0.24-(-0.52))$a_0$=0.76 $a_0$. To compare
theoretical predictions with the experiment we need only to know
the $a(|\vec B|)$ dependence around the zero crossing, which is
known at the percent level. This is $a(|\vec B|)=a_{bg} \cdot
(|\vec B|-B_{zc})/\Delta$, where $a_{bg}$ is the background
scattering length of K and $\Delta$ the width of the Feshbach
resonance we employ \cite{Fattori}. Using the measured values
given above, we can calculate ($B_{\parallel} - B_{\perp}$)
$\cdot$ $a_{bg}/\Delta$ and find $(0.36 \pm 0.1) a_0$, clearly not
in agreement with the prediction above that takes into account
only the on-site MDI. Solving the complete Eq. \eqref{DNLSE} we
find instead that the contribution of the inter-site dipolar
coupling is definitely not negligible, and the minima of
decoherence are achieved for $a$=-0.32 $a_0$ and $a$=0.11 $a_0$
for the OL $\parallel \vec B$ and the OL $\perp \vec B$
respectively, with a consequent separation of $0.43$ $a_0$.  The
agreement with the experiment is now much better, showing the
necessity of including the iter-site MDI in the model.

To get a deeper insight into the role played by the long-range
character of the dipolar interaction, we plot in Fig. 3a the
values of the on-site MDI $NU^{dd}_{j, j} |\psi_j|^2$ and the
values of the inter-sites MDI $N\sum_{\delta\neq 0}U^{dd}_{j,
j+\delta} |\psi_{j+\delta}|^2$ for the configuration OL $\parallel
\vec B$. The inter-site MDI is attractive, in agreement with the
qualitative analysis of Eq.\ref{dipolar pot} above, and of the
same order of magnitude as the on-site MDI. In Fig. 3 b) we plot
the total interaction energy for three cases: $a = 0$, where the
residual energy is the total MDI energy; $a=\bar a =-0.52$ $a_0$,
where the on-site MDI is perfectly canceled, and the residual
energy is due to the inter-sites MDI; $a= -0.32$ $a_0$, i.e. the
value that minimizes the decoherence. Note how a perfect
cancelation of the interaction energy is not possible due to the
different profiles of the curves in Fig. 3 a), and that the
minimum of the decoherence is achieved not when the total
interaction energy is averaged to zero, but when its variance is
minimized. The partial compensation of the dipolar interaction
with the contact interaction allows a reduction of the decoherence
rate of our alkali-based interferometer. The model predicts a
decoherence rate of 1 Hz for $a$=0 and a residual rate of 0.05 Hz
on the minima due to the uncompensated dipolar interaction.
Unfortunately we cannot test this prediction because technical
noise in our apparatus is presently one order of magnitude larger
than this. We plan to study this fundamental limit to the
interferometer's coherence with an optimized apparatus in the near
future. A higher sensitivity to interaction-induced decoherence
would also allow the verification of the presence of second order
effects that cannot be taken into account by our simple model such
as dipolar-induced dynamical instabilities \cite{Demler, Fallani}.
Note that our model predicts the possibility of completely
canceling the dipolar interaction by choosing a "magic angle"
$\theta=54.7^{\circ}$ between the dipoles and the lattice axis for
which $\xi=0$ \cite{Giovanazzi}. A comparison of the differential
measurement we have performed with the theoretical prediction can
also be used to determine with better accuracy the magnetic-field
position of the zero-crossing as $B_{zc}=(350.4 \pm 0.1)$ G. This
value is however in perfect agreement with the previous
determination by Feshbach spectroscopy \cite{FSK39}.

In conclusion, we have detected and studied the role of the
magnetic dipolar interaction (MDI) in a BEC-based atom
interferometer. We have shown that MDI-induced decoherence can be
suppressed by a proper choice of the scattering length. We have
proved that the interferometer is sensitive to the MDI between
different lattice sites. Our work constitutes a further step
towards the realization of a high sensitivity interferometer
employing a BEC with tunable interactions.

We acknowledge discussions with A. Simoni, M. Modugno, and the
rest of the quantum gases group at LENS. This work was supported
by MIUR (PRIN 2006), by EU(MEIF-CT-2004-009939), by INFN, and by
Ente CRF, Firenze. B. D. acknowledge support under ESA contract
SAI 20578/07/NL/UJ.


\begin{thebibliography}{99}
\bibitem{DipoloEff} Due to the hyperfine structure of $^{39}$K, at 350 G the internal state of the atom
in the $|I_z, S_z\rangle$ base, where $I$ is the nuclear spin and
$S$ the electronic spin, is $|\psi\rangle = \alpha |3/2,
-1/2\rangle + \beta |1/2, 1/2 \rangle$ with $\alpha =-0.987305$
and $\beta = 0.158831$. This means that the effective magnetic
dipole moment is $\langle \psi|\vec \mu|\psi \rangle =\langle
\psi| -2 \mu_B \vec S/\hbar |\psi \rangle = (\alpha^2 - \beta^2)
\mu_B \hat z = 0.95 \mu_B \hat z$.
\end{thebibliography}
\end{document}